\documentclass{emulateapj}
\shorttitle{Jets from quasi stars}

\shortauthors{Czerny et al.}

\begin{document}

\title{Quasi-star jets as  unidentified gamma-ray sources}
\author{Bozena Czerny$^{1}$, 
Agnieszka Janiuk$^{2}$, 
Marek Sikora$^{1}$}\affil{$^{1}$
Copernicus Astronomical Center, Polish Academy of Sciences, ul. Bartycka 18, 00-716 Warsaw, Poland}\email{bcz@camk.edu.pl; sikora@camk.edu.pl}
\affil{$^{2}$Center for Theoretical Physics, Polish Academy of Sciences, Al. Lotnikow32/46, 02-668 Warsaw, Poland}\email{agnes@cft.edu.pl}
\and
\author{Jean-Pierre Lasota$^{3,4}$}
\affil{$^{3}$ Institut d'Astrophysique de Paris, UMR 7095 CNRS, UPMC Univ Paris 06, 98bis Boulevard Arago, 75014 Paris, France}
\affil{$^{4}$ Astronomical Observatory, Jagiellonian University, ul. Orla 171, 30-244 Krakow, Poland}
\email{lasota@iap.fr}

\begin{abstract}
Gamma-ray catalogs contain a considerable amount of unidentified sources. Many of these are located out of the Galactic plane and therefore may have extragalactic origin.  Here we assume that the formation of massive black holes  in galactic nuclei proceeds through a quasi-star stage and  consider the possibility of jet production by such objects. Those jets would be the sources of collimated synchrotron and Compton emission, extending from radio to gamma rays. The expected lifetimes of quasi-stars are of the order of million of years while the jet luminosities, somewhat smaller than that of quasar jets, are sufficient to account for the unidentified gamma-ray sources. The jet emission dominates over the thermal emission of a quasi-star in all energy bands, except when the jet is not directed towards an observer. The predicted synchrotron emission peaks in the IR band,  with the flux close to the limits of the available IR all sky surveys.  The ratio of the $\gamma$-ray flux to the IR flux is found to be very large ($\sim 60$), much larger than in BL Lac objects but reached by some radio-loud quasars. On the other hand, radio-loud quasars show broad emission lines while no such lines are expected from quasi-stars. Therefore the differentiation between various scenarios accounting for the unidentified gamma-ray sources will be possible at the basis of the photometry and spectroscopy of the IR/optical counterparts.
\end{abstract}


\section{Introduction}

A significant number of sources in the $\gamma$-ray sky has no counterparts at other energy bands \citep{casandijan, abdo, bird} despite considerable efforts to identify them \citep{maeda, masetti}. This puzzle has persisted in gamma-ray astronomy since many years. Even the source catalog of the Fermi-LAT instrument with the best position errors contains about 300 of  still unclassified, high-latitude sources \citep{abdo}. The nature of these sources is thus difficult to establish but because of their positions they are most likely of extragalactic origin.  This means that  explaining  their nature requires the identification either with known active galaxies or the existence of a large population of a new type of extragalactic sources. The first option has been recently considered by Massaro et al. (2012a; 2012b). We address here the second possibility.

The part of the galaxy evolution which still remains the most mysterious is the pre-quasar epoch of  formation of a massive black hole. The process must be very rapid since massive black holes at galactic centers must form surprisingly early in order to account for the quasar-type activity at very large redshifts \citep{volonteri2006, haiman}.  More and more distant quasars located at very high redshifts are continuously being discovered, and their central black hole masses are large (e.g.  ULAS~J112001.48+064124.3: z = 7.085, $M = 2 \times 10^{9} M_{\odot}$; \citep{mortlock}). 

An attractive scenario for the growth of black holes in time considerably shorter than the Salpeter timescale has been proposed by \citet{begelman2006}. There the early accretion proceeds through a quasi-star stage, with a massive accreting envelope surrounding an initially small, few solar mass seed black hole. The detailed structure and quasi-stationary evolution of a quasi-star, assuming spherical symmetry, has been recently discussed in more detail by \citet{begelman2010}, \citet{volonteri2010}, \citet{ball}, and  \citet{ball2}. However, as in the case of a collapsing star, the configuration does not have to be spherically symmetric. If the quasi-star material posseses some angular momentum, a jet may form, as in  gamma-ray bursts or active galactic nuclei. A strong collimation of radiation may lead to enhancement of the observed radiation, large enough to make 
$\gamma$-rays detectable from cosmological distances and to make domination
of IR/optical synchrotron radiation over  uncollimated thermal radiation from 
the quasi-star's envelope. Finding such IR/optical counterparts to the 
$\gamma$-ray sources will provide oportunity to justify the model.
In the present paper we consider the possibility of the quasi-star jet formation and its observational consequences. 

\section{Model of a quasi star}

The formation of seeds for the present-day massive black holes is unclear and a number of possibilities are under discussion since the seminal paper by \citet{rees}, and various mechanisms may actually coexist. At large redshifts ($z \ge 20$) Population III stars likely form \citep{bromm} and finish their evolution as black holes with masses of order of a few hundred solar masses \citep{heger}. At intermediate redshifts, where young galaxies form,  protogalaxy mergers lead to creation of a strong gas inflow toward the center and a massive quasi-star may form. Such a spherical configuration is unstable but if partially supported by rotation, may form a quasi-stationary configuration of a quasi-star. A black hole at the center is accreting at highly super-Eddington rate and providing the energy support for the expanding self-gravitating envelope; the structure is therefore similar to that of a red giant star but with much higher mass. The inner black hole slowly grows till it contains about 1-10 per cent of the initial quasi-star's mass when the configuration looses stability and becomes dispersed, leaving behind an intermediate mass black hole. Such a scenario was proposed by \citet{begelman2006} and subsequently developed in a number of papers \citep{beg08, volonteri2010, ball}.

In the present paper we supplement the quasi-star scenario with the possibility of a jet formation which will shape its observational appearance. 
In this case the jet production is presumably similar to that occurring in AGN and GRBs, i.e.  mediated by rotation of poloidal magnetic fields in the BH and/or disk magnetosphere. Such fields can effectively be transported from the outer regions to the center only by geometrically thick accretion flows \citep{cao, mckinney, sasha}. 

Here we use a very simple parametric description of a quasi-star and the jet 
and study the observational consequences of this model. 

\subsection{The quasi-star structure}

\begin{figure}
\plotone{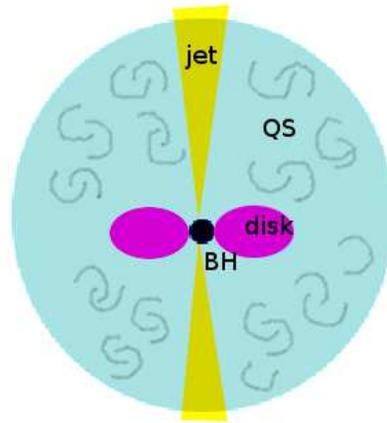}
\caption{The structure of a quasistar. Representative parameters: mass $\sim 10^{7} M_{\odot}$, luminosity $\sim 10^{45}$ erg s$^{-1}$, black hole mass $10^{5} M_{\odot}$, and the geometrically thick, advective disk with accretion rate $\sim 2 M_{\odot}$yr$^{-1}$, with radiative efficiency of 0.01. Accretion energy is converted to power the bipolar jets with an efficiency of 0.1 and the jets have Lorentz factor $\Gamma \sim 15$, radiative efficiency 0.1 and an isotropic luminosity $\sim 10^{48}$ erg s$^{-1}$. The jets produce gamma rays in the reconfinement shocks formed within 0.01-1 $r_{QS}$, i.e. $10^{15}-10^{17}$ cm.}
\label{fig:quasistar}
\end{figure}

A quasi-star (hereafter QS) in our scenario consists of two parts: a spherically symmetric envelope radiating thermally at the Eddington luminosity and a relativistic jet. Such a structure can be described using very few parameters. 
The schematic view of our model is shown in Fig. \ref{fig:quasistar}.

\subsection{The spherical component}

The basic parameter of the object is the quasi-star mass, $M_{\rm QS}$. We adopt as representative the value of $10^7 M_{\odot}$ and further use the total mass in the dimensionless units of$M_{\rm QS,7}$.

The QS isotropic thermal luminosity is simply determined by the condition of the Eddington luminosity at the surface of the spherical envelope,
\begin{equation} 
L_{\rm QS} = 1.3 \times 10^{45}\, M_{\rm QS,7}\, ~ [{\rm erg} \ {\rm s^{-1}}]. 
\end{equation}
The value of the central black hole mass, the surface temperature and the radius of the spherical configuration evolve. The mass of the central black hole grows, the envelope expands and the effective temperature drops. In general, the temperature is determined by the QS mass and the central black hole mass through a simple formula \citep[see Eq. 4 in][]{volonteri2010} which results from the polytropic approximation to the stellar structure, and the configuration is described by two parameters. In order to simplify the parameterization we use the final stage of the QS as representative. This final stage is achieved when the star reaches the Hayashi limit, roughly given by the effective temperature  $T_{\rm QS,min}$, equal to 3000 K for a solar metallicity. We can then determine the black hole mass of such configuration from Eq. 4 in \citet{volonteri2010},
\begin{equation}
M_{\rm BH} = 8.5 \times 10^5 \, M_{\rm QS,7}^{7/8} \, \epsilon_{\rm d,-2}^{-1/2} 
\,  ~[M_{\odot}]
\end{equation}
where $\epsilon_{\rm d}$ is the radiative efficiency of the disk. We expect this efficiency to be low because the accretion rate onto the central black hole is highly super-Eddington and the flow is advection-dominated, so we take the value of 0.01 as representative. The corresponding accretion rate ($ L_{\rm QS} = \epsilon_{\rm d} \, \dot M_{\rm d} \, c^2$) is
\begin{equation}
\label{eq:mdot}
\dot M_{\rm d} = {L_{\rm QS} \over \epsilon_{\rm d} c^2} 
= 2.2  \, M_{\rm QS,7}\,  \epsilon_{\rm d,-2}^{-1}\, ~ [M_{\odot}\,  {\rm yr^{-1}}]
\end{equation}
i.e. the Eddington ratio is of order of 100 at the end of the QS evolution, and much higher at earlier stages.

The thermal emission of the QS should be visible in the IR, in 2 to 3 micron wavelength range (K and L bands), taking into account the red color of the envelope, and the redshift. The assumption of $T_{\rm QS,min} = 3000 K$ determines also the radius of the QS
\begin{equation}
\label{eq:radius}
r_{\rm QS} = \sqrt{L_{\rm QS} \over 4\pi \sigma T_{\rm QS,min}^4} \simeq 
1.5 \times 10^{17} \, M_{\rm QS,7}^{1/2} \, ~[{\rm cm}],
\end{equation}
 where $\sigma$ is Stephan-Boltzman constant. The object is very extended due to the rapid decline in the density with radius.

The configuration is long-living:
\begin{equation}
\label{eq:t_life}
t_{\rm QS} \sim {M_{\rm BH} \over \dot M_{\rm d}} = 
3.8 \times 10^5 \, M_{\rm QS,7}^{-1/8} \, \epsilon_{\rm d,-2}^{1/2} \,~[{\rm yr}]
\end{equation}
although it lasts less than the quasar activity period. 

\subsection{The jet}

We further assume that a fraction of energy of the accreting material is converted to power the relativistic jet with an efficiency of $\eta_{\rm j}$. For a super-Eddington accretion flow the radiative efficiency $\epsilon_{\rm d}$ is diminished by advection but there are no such constraints for a jet, and  $\eta_{\rm j}$ can be higher than $\epsilon_{\rm d}$ so we take the value of 0.1 as a scaling factor. Therefore the total power of a quasi-star jet can be parameterized as 
\begin{eqnarray}
P_{\rm j} & = & \eta_{\rm j} \, \dot M_{\rm d} c^2 = {\eta_{\rm j} \over \epsilon_{\rm d}} L_{\rm QS}\nonumber \\
& = & 1.3 \times 10^{46} 
\, M_{\rm QS,7} \, {\eta_{\rm j,-1} \over \epsilon_{\rm d,-2}} \ \ [{\rm erg} \ {\rm s^{-1}}].
\end{eqnarray}

 For such a jet power the breakout condition,
$t_{QS} \gg r_{QS}/v_h$, is very well satisfied, where $v_h$ is the velocity 
of the jet head paving the way outwards through the QS and determined
by the balance between momenta fluxes of a jet and external medium as measured
in the head comoving frame (see, e.g. Scheuer 1974; Waxman \& M\'esz\'aros 2003). The head velocity is relatively slow when emerging from the star,  $v_h \sim 430$ km s$^{-1}$, but it is still highly supersonic due to the low temperature of the envelope.

If the radiative efficiency of the jet is $\epsilon_{\rm j}$ and its radiation is collimated into a solid angle $\Omega_{\rm rad} = \pi \theta_{\rm rad}^2$, then an observer located within this angle will detect a flux $F = {L_{\rm j,iso} / 4 \pi d_{\rm L}^2}$, where the isotropic jet luminosity is 
\begin{eqnarray}
\label{eq:lum_iso}
L_{\rm j,iso}& = & {4 \pi \over \Omega_{\rm rad}}\,  \epsilon_{\rm j} \, P_{\rm j} \nonumber \\
& = &1.1 \times 10^{48} \, 
\frac {\epsilon_{\rm j,-1} \eta_{\rm j,-1} \, M_{\rm QS,7}}
{\epsilon_{\rm d,-2} \, (\theta_{\rm rad} / 4^{\circ})^2} \, \ \
[ {\rm erg} \  {\rm s^{-1}}].
\end{eqnarray}
Here $4^{\circ}$ is the maximum angle at which the condition of the 
causal connection between the two opposite sides of a jet is satisfied for a jet Lorentz factor 
$\Gamma \simeq 15$ -- a typical value for jets in blazars associated with quasars.
Hence, the observed flux from the jet is by a factor
\begin{equation}  
{L_{\rm j,iso} \over L_{\rm QS}} = 8.5 \times 10^2 \, \epsilon_{\rm j,-1}\eta_{\rm j,-1}\epsilon_{\rm d,-2}^{-1} (\theta_{\rm rad}/4^{\circ})^{-2} \,
\end{equation}
larger than the flux from the QS. The observed jet emission always dominates over the thermal emission of the spherical envelope if the jet is directed towards an observer.

Accretion disks formed in the center of QSs are super-Eddington (see Eq.~\ref{eq:mdot} and the discussion below) and therefore geometrically thick. Such  disks can efficiently advect poloidal magnetic fields to the central regions of the accretion flow \citep{cao}. This may lead to formation of powerful, Poynting flux dominated jets \citep{mckinney, sasha}. Such jets undergo acceleration up to the moment when about half of their electro-magnetic power is converted to the kinetic energy of the carried matter, i.e. when $\sigma_{\rm B} = L_{\rm B}/L_{\rm K}$ drops to unity, where $L_{\rm B}$ is the magnetic energy flux, $L_{\rm K}$ is the kinetic energy flux, and $P_{\rm j}= L_{\rm B} + L_{\rm K}$. This happens at a distance $r \sim 10^2 - 10^3 r_{\rm g}$, where $r_{\rm g}= GM_{\rm BH}/c^2$ \citep{komis, tchek, lyub}. Further away the conversion proceeds much slower, but possibly at $r \sim 10^4 - 10^5 r_{\rm g}$ the ratio $\sigma_{B}$ becomes small enough to allow for the formation of strong shocks and efficient particle acceleration \citep{sik}. Because the size of our QS at $t=t_{\rm QS}$ is given by Eq.~\ref{eq:radius} while the gravitational radius of a black hole with the mass given by Eq.~2 is
\begin{equation}
r_{\rm g} = 1.3 \times 10^{11} M_{\rm QS,7}^{7/8} \epsilon_{\rm d,-2}^{-1/2} \, [{\rm cm}] \, 
\end{equation}
we have a very large ratio of the envelope to the black hole radii $r_{\rm QS}/r_{\rm g} \sim 10^6$. Therefore one may expect that efficient shock formation and non-thermal radiation production takes place within a distance range $0.01 - 1 ~r_{\rm QS}$, where due to the interaction with the QS's envelope the reconfinement shocks develop. 

\section{Radiative properties}

We now estimate the properties of radiation emerging from a QS with a jet in order to formulate the observational characteristics of such objects.

\subsection{Radiative processes}

The dominant radiative  processes are: synchrotron, synchrotron-self-Compton (SSC), and external-radiation-Compton (ERC). Radiative electron energy losses in these processes scale with the magnetic energy density, $u_{\rm B}'$, synchrotron energy density, $u_{\rm syn}'$, and the external radiation energy density, $u_{\rm ext}'$, respectively, all measured in the jet co-moving frame.
Noting that $L_{\rm B} \equiv c u_{\rm B}' \pi r^2 (\theta_{\rm j} \Gamma)^2 $ and
that for $\sigma_{\rm B} \ll 1$, $L_{\rm B}=\sigma_{\rm B} P_{\rm j}$, we have 
\begin{equation}
\label{eq:u_prim_B}
u_{\rm B}'= {\sigma_{\rm B} P_{\rm j} \over c \pi r^2 (\theta_{\rm j} \Gamma)^2} \, 
\end{equation}
and
\begin{equation}
u_{\rm ext}' \simeq \Gamma^2 u_{\rm ext} = 
{4 \over 3} {\tau L_{\rm QS} \Gamma^2 \over 4 \pi r^2 c} \, ,
\end{equation}
we find the ratio of the external to the synchrotron flux equal to
\begin{equation}
{F_{\rm ERC} \over F_{\rm syn}} = {L_{\rm ERC} \over L_{\rm syn}} \simeq 
77 {\epsilon_{\rm d,-2}\tau (\Gamma/15)^2 (\theta_{\rm j} \Gamma)^2 \over 
\sigma_{\rm B,-1} \eta_{\rm j,-1}} \, ,
\end{equation}
where $\theta_{\rm j}$ is the opening angle of a jet.  The parameter $\tau$ scales the real density of QS radiation entering a jet at a distance r by a factor $L_{\rm QS}/4 \pi r^2 c$. At $r=r_{\rm QS}$ it is $\tau \sim 1$, and at $r< r_{\rm QS}$ it is $\tau >1$. Its dependence on a distance within the QS envelope  is determined by the dependence of the opacity  on radius as well as on multi-scatterings of QS photons off the jet walls. Making the comparison of $u_{\rm B}'$ with $u_{\rm syn}'$ one can find that  $L_{\rm syn} \gg L_{\rm SSC}$. 

\subsection{Radiation spectra}

In the presence of protons, the electrons are energized in at least two steps, the preheating and the stochastic. In the first step, due to coupling via collective processes in magnetized plasma with protons, electrons are preheated up to thermal energies of the shocked protons \citep{spit, sironi}. In the second step, electrons together with protons participate in the Fermi 1 acceleration process \citep{be, no}. For relativistic shocks (Lorentz factor of the upstream flow relative to the shock front is $ > 2$), the electrons reach ultra-relativistic energies already in the first step. This energy in $m_{\rm e}c^2$ units is $ \gamma_{\rm m} = \kappa m_{\rm p}/m_{\rm e}$, where for mildly relativistic shocks $\kappa \sim 1$. Electrons with this energy are responsible for the luminosity peak in the $\nu L_{\nu}$ representation.

Electrons with such energies produce ERC spectra peaking at photon energies
\begin{equation}
{\rm h} \nu_{\rm ERC} \simeq (\Gamma \gamma_{\rm m})^2 \, {\rm h}\nu_{\rm ext}  (1+z)^{-1} 
\, , 
\end{equation}
 where ${\rm h}$ is Planck constant and $z$ is the source cosmological redshift. 

Noting that the peak of the thermal spectrum in the $\nu L_{\nu}$ representation
is $\nu_{\rm ext} \simeq 3.9 ~kT$, and that 
\begin{equation}
T = 3000 \, \tau^{1/4} \, (r_{\rm QS}/r)^{1/2} \ \ [{\rm K}] \, ,
\end{equation}
one finds
\begin{equation}
{\rm h}\nu_{\rm ERC} \sim 100 (\Gamma/15)^2 \, (r_{\rm QS}/r)^{1/2} \tau^{1/4} \,\, [{\rm MeV}] \,  .
\end{equation}
Hence for any location of radiative processes within a distance range
$0.01 - 1 \ \ r_{\rm QS}$, the dominant radiative output is predicted to be within
the Fermi energy band.

The $\gamma_{\rm m}$ electrons will produce synchrotron spectra at
$\nu_{\rm syn} \simeq 3.7 \times 10^6 \gamma_{\rm m}^2 B' \Gamma /(1+z)$ Hz, and for 
$u_{B}'$ given by Eq.~\ref{eq:u_prim_B}
\begin{equation}
B'= 3.9 \,  \sigma_{B,-1}^{1/2} \,\left( r_{QS}/r \right) \, 
(\theta_j\Gamma)^{-1} \eta_{j,-1}^{1/2} \epsilon_{d,-2}^{-1/2} \,\, [{\rm G}] \, 
,
\end{equation}
the peak of the synchrotron radiation produced at $z=3$ is located at 
frequencies
\begin{equation}
\nu_{\rm syn} \simeq 2.0 \times 10^{14} \, \frac {(\Gamma/15) \kappa^2 
\sigma_{\rm B,-1}^{1/2} \, (r_{\rm QS}/r) \, \eta_{\rm j,-1}^{1/2}}
{(\theta_j\Gamma) \epsilon_{d,-2}^{1/2}}
\,\, [{\rm Hz}] \, .   
\end{equation}

Our results show that QS jets produce synchrotron radiation in the IR band and that this radiation dominates over the QS thermal spectrum. Hence, the diagnosis based on IR spectral properties developed by \citet{dabrusco} might not distinguish between blazars and QSs. However, one may discriminate between these two classes of objects noting that we predict a ratio $L_{\rm Fermi}/L_{\rm IR} \sim$ several tens. This is observed in some luminous  blazars associated with quasars, but practically never in BL Lac objects. On the other hand we can discriminate QSs from quasar-associated blazars by noting that in the latter the broad emission lines are seen, whereas they are not expected in QSs. 

\subsection{Statistics}

The collimation of the jet radiation increases the observed luminosity but the high collimation of the quasi-star jet lowers the probability of its detection. On the other hand, all galactic centers should go through this phase while only a small fraction of black holes pass through a stage of a bright quasar. Therefore statistically the number of sources is significant and should match the fraction of the still unidentified gamma ray sources. Below we consider this issue quantitatively. 

The energy flux for the weak unidentified Fermi sources is of the order of $10^{-11}$ erg cm$^{-2}$ s$^{-1}$ and such fluxes correspond to absolute luminosities of  $1.1 \times 10^{48}$ erg s$^{-1}$ (see eq.~\ref{eq:lum_iso}) at the redshift z = 3.4, in the standard cosmology (h = 71, $\Omega_{\rm m} = 0.27$, $\Omega_{\Lambda}= 0.73$). 

The number of quasi-stars forming at redshifts between 2 and 4 has been estimated to be of the order of 0.01  per cubic Mpc in the comoving frame by \citet{volonteri2010}  (see their Fig. 5). With this rate we expect a total number of $ 10^{10}$ quasi-stars to form in this redshift range. The number of sources to be observed at a quasi-star stage will be much smaller since this stage lasts thousands to millions of years which is short compared to the age of the universe. A rough estimate of the QS phase duration is given by $t_{\rm QS}$ (see Eq.~\ref{eq:t_life}), thus reducing the number of active quasi-stars to a representative value of $3.1 \times 10^{5}$ for all dimensionless parameters equal 1. Since the emission is collimated, the number of sources with jets toward an observer will be further reduced. In general, combining all coefficients we obtain
\begin{equation}
N = 400 ( \theta/4^{\circ})^{2} M_{\rm QS,7}^{-1/8} \, \epsilon_{\rm d,-2}^{1/2}  \, ,
\end{equation}
which shows that both the typical $\gamma$-ray fluxes and the number of the high-latitude, unidentified Fermi sources in the First Fermi LAT Catalog \citep{abdo} can be reproduced in the QS-jet scenario.

Since the number of on-axis QSs at still larger redshifts (from 4 to 10) is only by a factor of 2 larger than the number of resolved sources, their contribution to the still unexplained $\gamma$-ray background \citep{inoue} can be significant (up to 50\%) taking into consideration their bolometric luminosity, but more careful analysis should be done taking into account the source and the background spectral shape. Off-axis sources can contribute to the hard X-ray background, but not significantly (a few per cent or less).

\section{Conclusions}

A large fraction of unidentified $\gamma$-ray sources is located at high latitudes and therefore most of them are considered to be extragalactic. We propose that quasi-star jets can account for these sources. In analogy to blazars they are predicted to produce nonthermal spectra, at lower energies (optical-IR) dominated by the synchrotron mechanism and in the $\gamma$-ray band by the inverse-Compton process. However, they can be distinguished from blazars by the following properties:

(i) the expected ratio of the gamma-ray to the IR 
components is close to 60, much larger than in BL Lac objects 
(see e.g. Fig. 8 of D'Abrusco et al. 2012)  but achievable by some luminous blazars associated with quasars; 

(ii) no broad emission lines are expected as in most of BL Lac objects but unlike in radio-loud quasars. 

Since the typical Fermi flux of the unidentified sources is of the order of  $10^{-11}$ erg cm$^{-2}$ s$^{-1}$, the predicted IR/optical  magnitudes of the QS 
counterpart are $\sim 17 - 18 $ mag in K band and $20 - 21 $ mag in R band. The currently available IR all sky survey WISE has the limiting magnitude of $16 - 17$,  which is at the border of the QS detection. Optical SDSS survey also covers a significant fraction of the sky and contains sources down to 22.5 mag in R band \citep{sloan8}. Since the IR/optical emission is non-thermal and likely highly variable, color diagrams and variability can help to select proper candidates.

\section*{Acknowledgments}

We are grateful to Mitch Begelman for helpful discussions. We acknowledge the support from the Polish Ministry of Science and Higher Education through the grants  NN 203 518 638, N N203 380336,NCN DEC-2011/01/B/ST9/04845, and from the French Space Agency CNES.

\end{document}